\documentclass[apjl]{emulateapj}

\usepackage{epsfig}
\usepackage{xspace}
\usepackage{color}

\newcommand{\rmn}{\mathrm}
\newcommand{\xvec}{{\bf{x}}}
\newcommand{\LAF}{\mathrm{Ly}\alpha}
\newcommand{\kb}{k_{\mathrm{o}}}
\newcommand{\bo}{b_{\mathrm{o}}}
\newcommand{\al}{\alpha}
\newcommand{\zre}{z_{\mathrm{RE}}({\bf{x}})}
\newcommand{\te}{\tau}
\newcommand{\xe}{x_{\mathrm{e}}}
\newcommand{\zbar}{\bar{z}}

\newcommand{\Delz}{\Delta_{\mathrm{z}}}
\newcommand{\Lik}{\mathcal{L}}

\slugcomment{Submitted to ApJ}

\shorttitle{CMB Polarization and Patchy kSZ power spectra}
\shortauthors{Battaglia et al.}

\begin{document}

\title{Reionization on Large Scales III: Predictions for Low-$\ell$ Cosmic Microwave Background Polarization and High-$\ell$ Kinetic Sunyaev-Zel'dovich Observables}

\author{N. Battaglia\altaffilmark{1}, A. Natarajan\altaffilmark{1}, H. Trac\altaffilmark{1}, R. Cen\altaffilmark{2}, A. Loeb\altaffilmark{3}}
\altaffiltext{1}{McWilliams Center for Cosmology, Wean Hall, Carnegie Mellon University, 5000 Forbes Ave., Pittsburgh PA 15213, USA}
\altaffiltext{2}{Department of Astrophysical Sciences, Princeton University, Princeton, NJ 08544}
\altaffiltext{3}{Harvard-Smithsonian Center for Astrophysics, Cambridge, MA 02138}

\begin{abstract}

We present new predictions for temperature (on small angular scales) and polarization (on large angular scales) CMB anisotropies induced during the epoch of reionization (EoR). Using a novel method calibrated from radiation-hydrodynamic simulations we model the EoR in large volumes (L $\gtrsim 2$ Gpc/$h$). We find that the EoR contribution to the kinetic Sunyaev- Zel'dovich power spectrum (patchy kSZ) ranges between $\sim$0.6 - 2.8 $\mu$K$^2$ at $\ell = 3000$, for the explored parameter space. For each model, the patchy kSZ power spectrum is calculated from 3 large $15^{\circ} \times 15^{\circ}$ maps for better numerical convergence. Decreasing the size of these maps biases the overall patchy kSZ power to higher values. We find that the amplitude of the patchy kSZ power spectrum at $\ell = 3000$ follows simple scalings of $D_{\ell=3000}^{\rmn{kSZ}} \propto \zbar$ and $D_{\ell=3000}^{\rmn{kSZ}} \propto \Delz^{0.47}$ for the mean redshift ($\zbar$) and duration ($\Delz$) of reionization. Using the constraints on $\zbar$ from the WMAP 7-year results and the lower limit on $\Delz$ from EDGES we find a lower limit of $\sim 0.4\,\mu$K$^2$ at $\ell = 3000$. Planck will constrain the mean redshift and the Thomson optical depth from the low-$\ell$ polarization power spectrum. Future measurements of the high-$\ell$ CMB power spectrum from the South Pole Telescope (SPT) and the Atacama Cosmology Telescope (ACT) should detect the patchy kSZ signal if the cross correlation between the cosmic infrared background and the thermal Sunyaev Zel'dovich effect is constrained. 
We show that the combination of temperature and polarization measurements constrains both $\zbar$ and $\Delz$. The patchy kSZ maps, power spectra templates and the polarization power spectra will be publicly available.

\end{abstract}


\section{Introduction}

Free electrons in the intergalactic medium (IGM) scatter photons from cosmic microwave background (CMB) creating additional secondary anisotropies that distort the primordial anisotropies. These free electrons are initially ionized from the neutral IGM by the first stars and galaxies during the epoch of reionization (EoR). Thus, information on EoR is imprinted on the CMB in both temperature and polarization. The temperature fluctuations are affected by the kinetic Sunyaev Zel'dovich (kSZ) effect, which is doppler shifting of CMB photons from the bulk motions of free electrons with respect to the CMB rest frame \citep{Suny1980}.The polarization signal on large angular scales is affected by the rotation sourced by free electrons from the beginning of EoR to the present \citep[e.g.][]{Bond1987,Hu1997,Zald1997,Liu2001,WMAP7cos}, which induces a curl free polarization signal (E-mode). These are the two leading order effects, while there are other smaller order effects, such as fluctuations in the optical depth \citep{ZR2}, that are not discussed.

Already measurements from the opacity of the $\LAF$ forest \citep{Fan2006a}, the redshifted 21 cm signal \citep{Bow2012} from the experiment EDGES\footnote{www.haystack.mit.edu/ast/arrays/Edges}, and the large-scale polarization of the CMB \citep{WMAP7pars} infer that reionization was extended. Further model dependent constraints on EoR come from: measurements of quasar proximity zones \citep[e.g.][]{Wyit2005,Fan2006b}, a null result for intergalactic damping wing absorption in a $z=6.3$ gamma-ray burst spectrum \citep[e.g.][]{Tota2006,McQn2008}, detections of damping wing absorption in the IGM from quasar spectra \citep[e.g.][]{Mess2004,Mess2008,Bolt2011}, and $\LAF$ emitter number densities and clustering measurements \citep[e.g.][]{Malh2004,Haim2005}. Recently, the South Pole Telescope (SPT) placed a model dependent upper limit on the duration of reionization \citep{Reic2012,Zahn2012} from their multifrequency measurements of the high-$\ell$ power spectrum of CMB secondary anisotropies. New CMB measurements of temperature and polarization anisotropies from the Planck satellite, the POLARBEAR experiment, the Atacama Cosmology Telescope (ACT), ACT-pol (ACT with polarization), SPT, SPT-pol (SPT with  polarization), and CMBpol \citep{Zald2008} have the potential to constrain the EoR from CMB measurements alone.

The amplitude of the EE power spectrum at $\ell \lesssim 20$ essentially measures the optical depth to reionization, $\te$, with the most recent constraints being $\te = 0.087 \pm 0.015$ and a mean reionization-redshift is $10.5\pm1.2$ (68\% CL). This amplitude is often predicted by CAMB \citep{Lewis2000}, or codes like it, using a parametric hyperbolic tangent function for the ionization history. There are modifications to CAMB \citep{Mort2008}, which allow any ionization history as an input.

The fractional contributions to the kSZ from the EoR \citep[e.g.][hereafter we refer to this contribution as the patchy kSZ]{Gruz1998,Knox1998,Vala2001,Sant2003,Zahn2005,McQn2005,iliv2007,Mess2011,Zahn2012,Mess2012} are the largest on small angular scales compared to the primary and other secondary CMB anisotropies. This patchy kSZ power is in addition to the kSZ power that comes from lower redshift \citep[][here after we refer to this contribution as the homogeneous kSZ]{Ostr1986,Jaff1998,Ma2002,Zhan2004,Shaw2012}. Many of the previous models for the patchy kSZ signal were calculated in small volumes ($\lesssim 1$ Gpc/$h$) and do not capture the large scale features of the patchy kSZ maps, which is required to accurately calculate the power spectrum.

This is the third paper (Paper III) in a series that explores EoR observables produced via our semi-analytical models of reionization that are statistically informed by simulations with radiative transfer and hydrodynamics. We introduce our model in Paper I \citep{ZR1}, we look at the impact of a patchy optical depth on CMB observables in Paper II \citep{ZR2}, and we explore the 21cm signal in Paper IV \citep{ZR4}.

We present in this paper predictions for CMB observables. These predictions are made in large volumes (L = 2 Gpc/$h$) and the importance of going to such large volumes is demonstrated throughout this work. The EE polarization power spectra, the kSZ power spectra, and the maps from this paper will be made publicly available. In Section~\ref{sec:mod}, we summarize our fast semi-analytical model and the simulations it is based on. In Section~\ref{sec:res}, we present results for the EE power spectrum and the kSZ power spectrum. We discuss prospects for future measurements and conclude in section~\ref{sec:con}. We adopt the concordance cosmological parameters that are consistent with WMAP 7-year results \citep{WMAP7pars}: $\Omega_{\rmn{m}} = 0.27$, $\Omega_{\Lambda} = 0.73$, $\Omega_{\rmn{b}} = 0.045$, $h = 0.7$, $n_{\rmn{s}}= 0.96$, and $\sigma_8  = 0.80$. 

\begin{figure*}
  \resizebox{0.5\hsize}{!}{\includegraphics{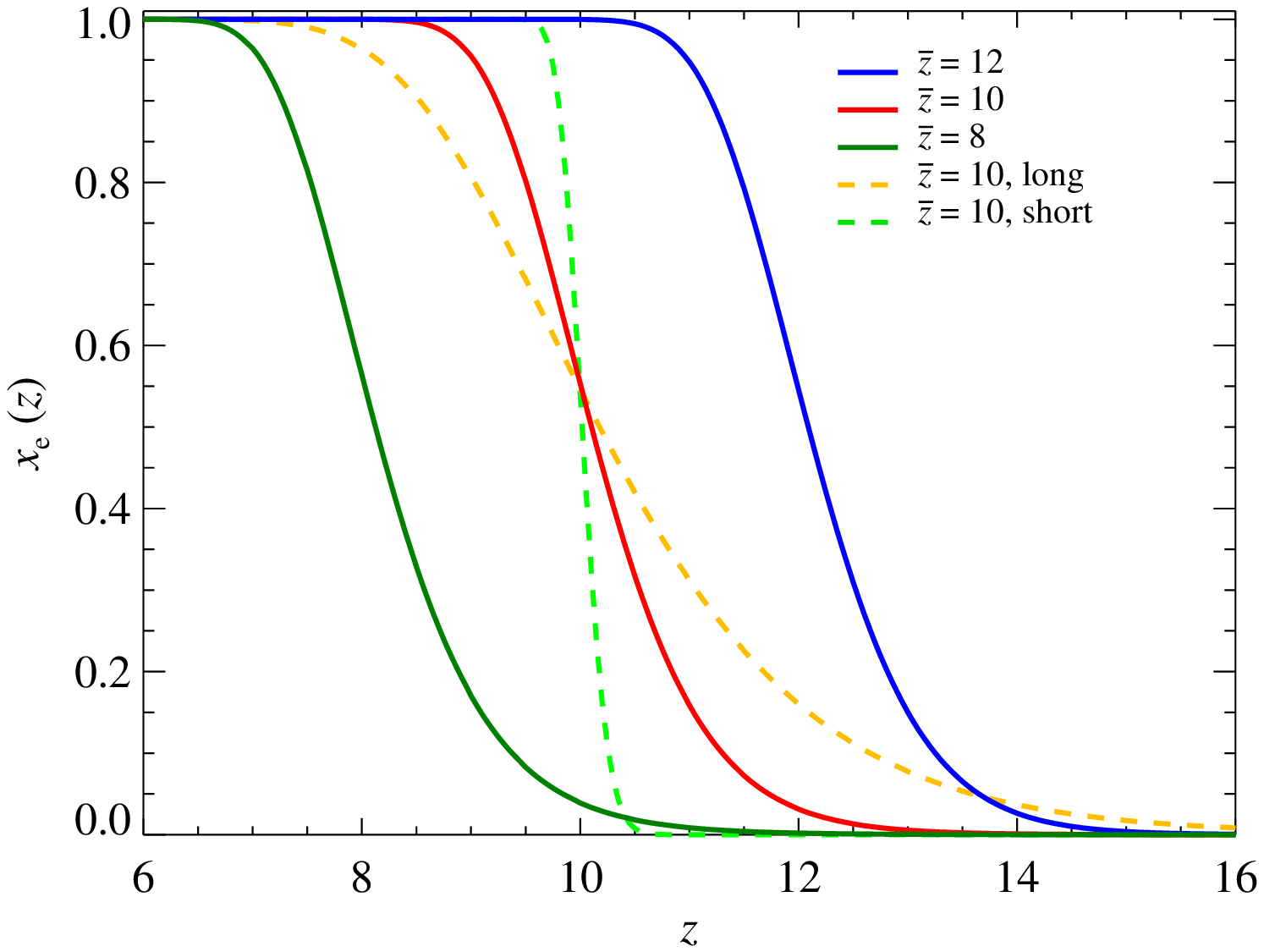}}%
  \resizebox{0.5\hsize}{!}{\includegraphics{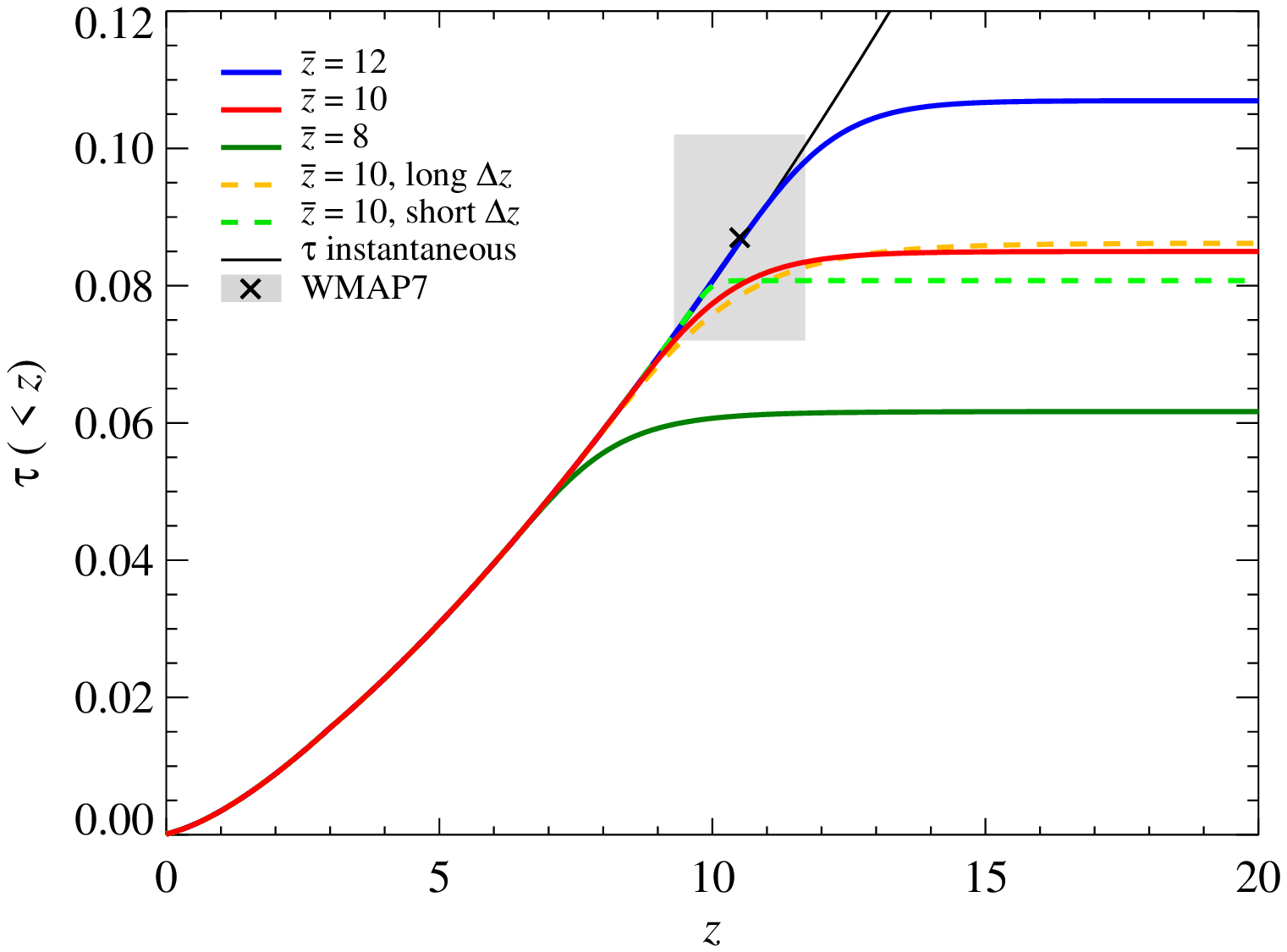}}\\                                                                                    
  \caption{Left: The ionization fraction as a function of redshift, $\xe(z)$, for five models, our fiducial model at at $\bar{z} =$ 8,10, and 12 (green, red, and blue, respectively) and two extreme models of brief (orange dashed) and long (light green dashed) duration reionization at a fixed $\bar{z}=10$. The $\xe(z)$ for the fiducial models have similar shapes and they are just shifted according to $\zbar$. Right: The corresponding optical depth, $\te$, for the same models as $\xe$. The values of $\te$ are compared against the WMAP 7-year constraints \citep[light grey box][]{WMAP7pars} on $\te$ and their reionizatoin-redshift. Like the polarization power spectrum, a constraint on $\te$ will differentiate between models with different $\bar{z}$, but cannot help differentiate between our models with large or small $\Delz$.}
\label{fig:tau}
\end{figure*}

\section{Parametric model for Reionization}
\label{sec:mod}

\begin{figure}
\epsscale{1.2}
\plotone{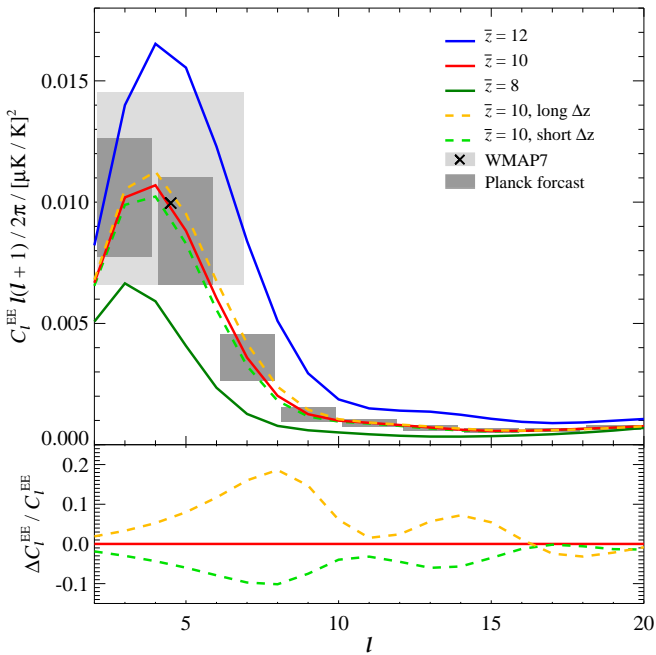}
\caption{Top: The low-$\ell$ EE polarization power spectrum for several models of reionization compared to the WMAP 7-year band power constraint \citep[light grey box][]{WMAP7pars} and the projected error bars from Planck \citep[dark grey bands][]{PlanckBB} for the 143 GHz channel calculated for our fiducial model. We show five models, our fiducial model at at $\bar{z} =$ 8,10, and 12 (green, red, and blue, respectively) and two extreme models of brief (orange dashed) and long (light green dashed) duration reionization at a fixed $\bar{z}=10$. Bottom: The percent difference between our fiducial and the two extreme models. The low $\ell$ EE polarization power spectrum will tightly constrain $\bar{z}$. However, it is not possible to discern between our models with large or small $\Delz$ for same $\bar{z}$ using only this measurement.}  
\label{fig:clee}
\end{figure}

In Paper I we developed a semi-analytic model for reionization based upon results from RadHydro simulations \citep[for more details see][]{Trac2008,ZR1}. In the simulations we construct a reionization-redshift field, $\zre$, that tracks the redshift at which each the gas cell becomes 90\% ionized. We define the following fluctuations fields for density

\begin{equation} 
\delta_{\rmn{m}} (\xvec) \equiv \frac{\rho(\xvec) - \bar{\rho}}{ \bar{\rho}},
\label{delm}
\end{equation}
\noindent and the reionization-redshift
\begin{equation} 
\delta_{\rmn{z}} (\xvec) \equiv \frac{[1 + \zre] - [1 + \zbar]}{1 + \zbar}
\label{delz}
\end{equation}

\noindent where $\bar{\rho}$ is the mean matter density and $\zbar$ is the mean value for the $\zre$ field,which is approximately equal to
the redshift of 50\% ionization. The fluctuations in both the $\delta_{\rmn{m}}$ and $\delta_{\rmn{z}}$ fields are highly correlated on scales $\gtrsim 1$ Mpc/$h$. We calculate a simple scale-dependent linear bias that relates these two fields and we represent this bias with the simple parametric form,

\begin{eqnarray}
b_{\rmn{zm}} (k) &=& \left[\frac{\left<\delta_{\rmn{z}}(k)\delta_{\rmn{z}}(k)\right>}{\left<\delta_{\rmn{m}}(k)\delta_{\rmn{m}}(k)\right>}\right]^{1/2} \nonumber \\
 &=& \frac{\bo}{\left(1 + k/\kb\right)^{\al}},
\label{eq:bias_form}
\end{eqnarray}

\noindent that contains 3 parameters  $\bo$, $\kb$, and $\al$. The value for $\bo$ that we use is determined from analytical arguments in \citet{Bark2004}. The fiducial parameter values for $\kb = 0.185$ Mpc/$h$ and $\al = 0.564$ are found by fitting the bias calculated from the simulations. We explore the parameter space of our model by varying $\kb$ and $\al$. The effects these parameters have on EoR are the following: increasing $\kb$ lengthens the duration of reionization while increasing $\al$ shortens reioinization \citep{ZR1}. Physically, a shorter reionization process tends to have larger ionization bubble sizes that percolate more quickly, which in turn correspond to more luminous ionizing sources.

\begin{figure*}
  \resizebox{0.33\hsize}{!}{\includegraphics{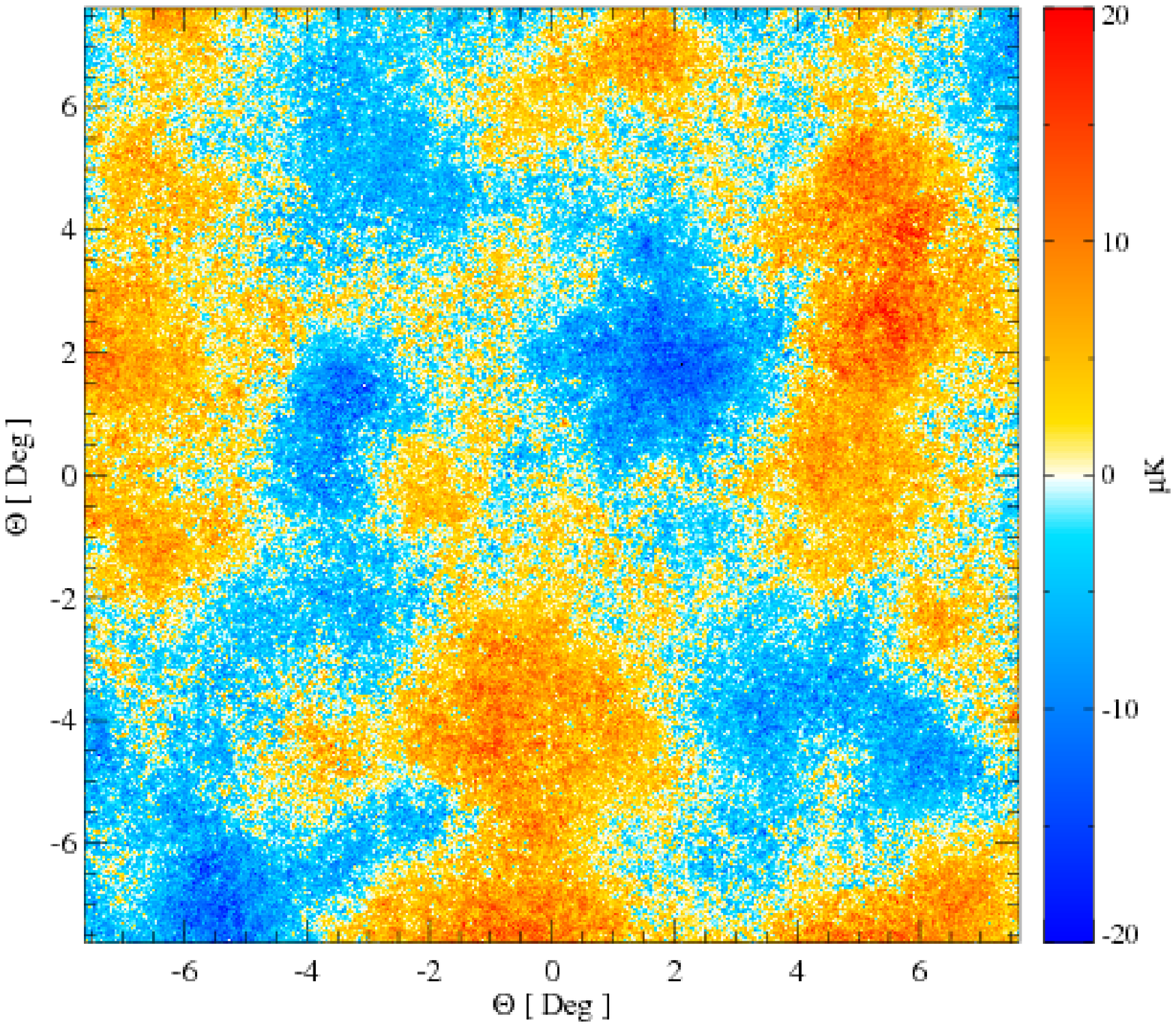}}%
  \resizebox{0.33\hsize}{!}{\includegraphics{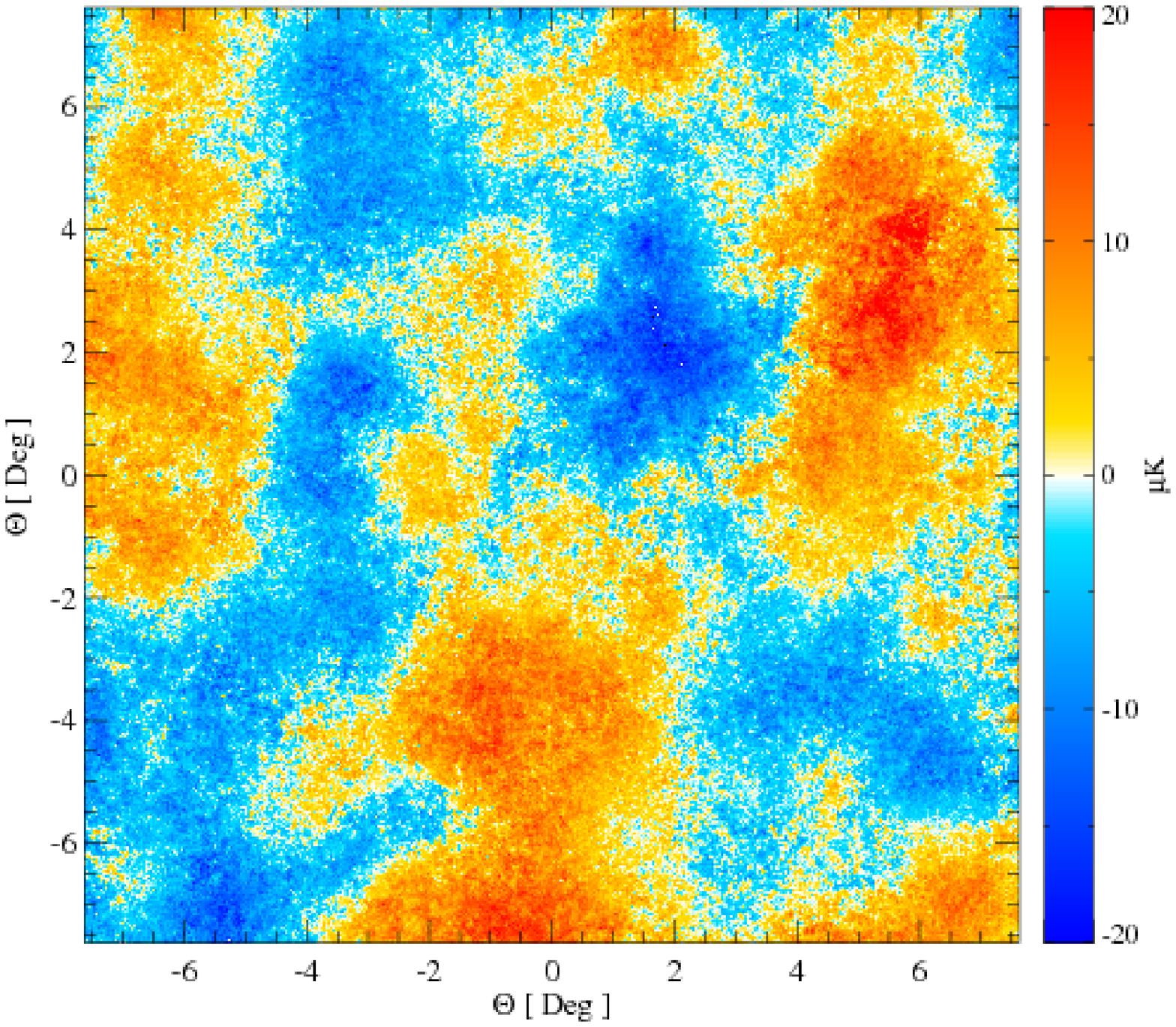}}%
  \resizebox{0.33\hsize}{!}{\includegraphics{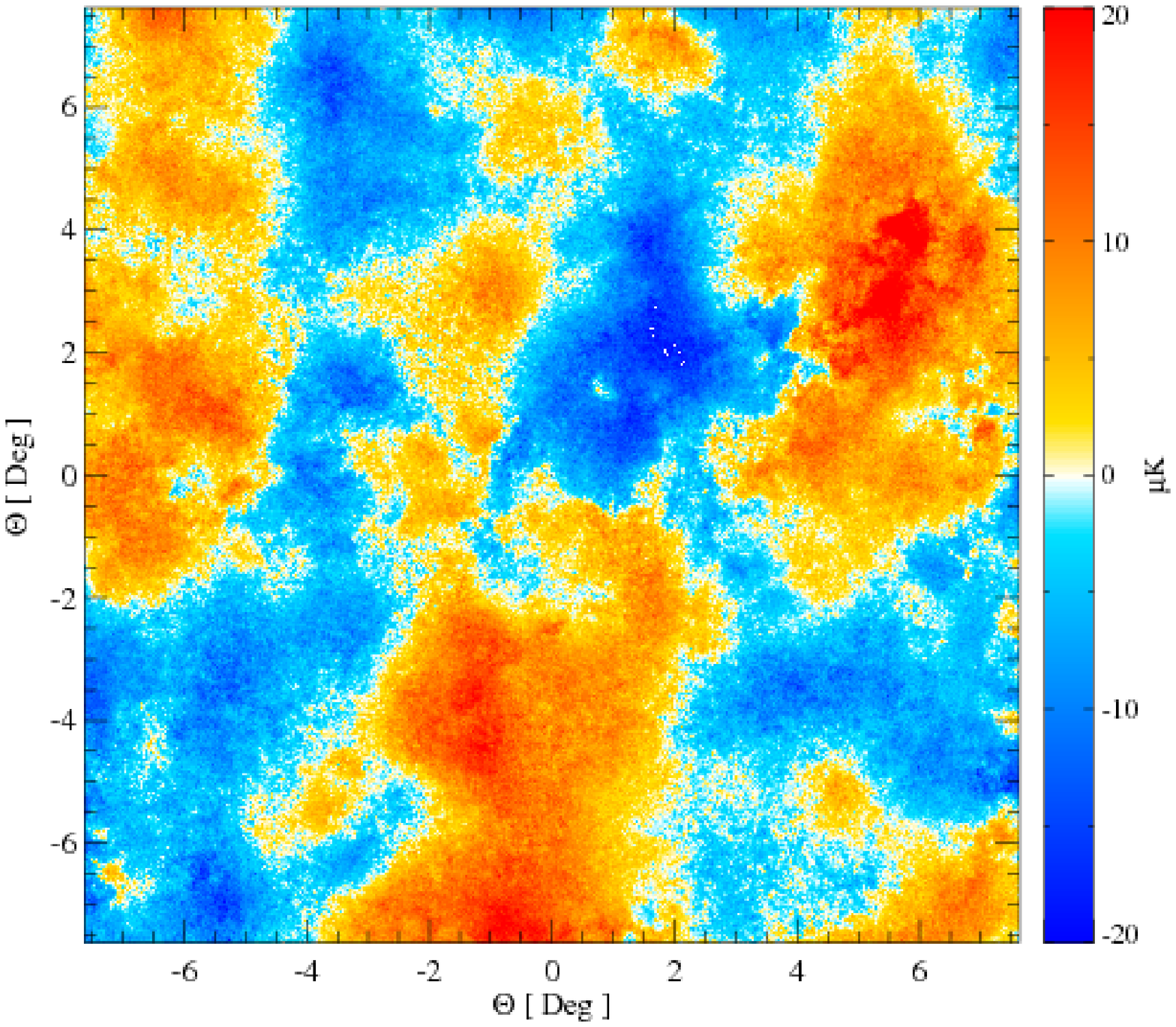}}\\                                                                                                      
  \caption{Light cone projected maps of the kSZ signal from patchy reionization for models with $\bar{z} = 10$ and our longest $\Delz$(left), our fiducial $\Delz$(center), and our shortest $\Delz$ (right). The overall large-scale structure is similar but the small-scale structure decreases as $\Delz$ decreases.}  
\label{fig:kszmap}
\end{figure*}

We generate the over-density fields, $\delta_{\rmn{m}}$ using a particle-particle-particle-mesh (P$^3$M) N-body code that evolves 2048$^3$ dark matter particles in a 2 Gpc/$h$ box down to $z= 5.5$. This over-density field is then convolved with a filter consisting of three elements: (1) a cubical top hat filter, $\Xi(k)$, which deconvolves the smoothing used to construct $\delta_{\rmn{m}}$ from the simulation, (2) a Fourier transform of a real space top hat filter $\Theta(k)$, which smoothes $\delta_{\rmn{m}}$ to a resolution of 1 Mpc/$h$, and (3) the bias function from Equation \ref{eq:bias_form}. The assembled filter takes this form 

\begin{equation}
W_{\rmn{z}}(k) = \frac{b_{\rmn{zm}} (k) \Theta(k)}{\Xi(k)},
\end{equation}

\noindent and we apply this filter at $\zbar$. The newly constructed $\delta_{\rmn{z}}$ field is Fourier transformed back to real space and converted to the $\zre$ field by Eq. \ref{delz}
with the same $\zbar$ as the density field. Here the value of $\zbar$ essentially sets the midpoint of reionization. We now have a complete ionization history for the density field used, which is then used to make ionization fields and kSZ maps. We define the duration of reionization as

\begin{equation} 
\Delz \equiv z(\xe = 25\%) - z(\xe = 75\%),
\label{Delz}
\end{equation}

\noindent where $\xe$ is the ionization history. This definition excludes the early and late times of reionization, since the small scale physical processes at these times are difficult to capture, hence is insensitive to the epochs when both our simulations and semi-analytic method are most uncertain. For a detailed parameter study of $\Delz(\kb,\al)$ see \citet{ZR1}.

\section{Results}
\label{sec:res}
We present our model predictions for the integrated optical depth, $\te$, the low-$\ell$ EE mode polarization power spectrum, and the contribution to the patchy kSZ power spectrum. These predictions are compared to current and projected constraints, as well as previous work.
 
\subsection{Optical Depth and EE Polarization Power Spectrum} 

We calculate the large scale EE polarization power spectrum using CAMB \citep{Lewis2000} with the modifications by \citet{Mort2008}, which accepts general reionization histories, $\xe(z)$. Although novel, we did not include their PCA implementation. Figure \ref{fig:tau} shows the results for $\xe(z)$ from our fiducial models at $\zbar = 8,10,12$ and two extreme models for long and short duration reionization scenarios at $\zbar = 10$.
We show the corresponding EE polarization power spectra in Figure \ref{fig:clee} and compare them to the WMAP 7-year band power constraints \citep{WMAP7pars}. We find the fiducial parameters with $\bar{z} = 10$ agree with the WMAP constraint, and the 1-$\sigma$ confidence interval is bracketed by choices of $\bar{z} = 8$ and  $\bar{z} = 12$. Our two extreme models of short and long duration reionization (here after  small and large $\Delz$) at fixed $\bar{z}$ differ from the fiducial model by a maximum 20\% (cf. Fig \ref{fig:clee}). Illustrated in Fig. \ref{fig:clee} is that the current WMAP 7-year data is unable to differentiate between these extreme models for reionization and given the projected error bars on the EE power spectrum \citep[][for the 143 GHz channel and our fiducial model]{PlanckBB} neither will the upcoming observations from Planck. Thus, the low $\ell$ EE polarization power spectrum constraints are not sensitive to the duration of reionization. This conclusion is the same as previous work by \citet{Zahn2012}, however, they come to this conclusion via a different semi-analytic model. Similarly, WMAP showed that only using primary CMB constraints there is a degeneracy between the duration and the mean redshift of reionization \citep{WMAP7cos}. There is no benefit to include the low $\ell$ TE cross spectrum, since the projected error bars from Planck \citep{PlanckBB} are too large to differentiate between small and large $\Delz$ models.

Any constraint on $\te$ is primarily driven by the measurement of the low $\ell$ EE polarization power spectrum.  To first order the amplitude of the low $\ell$ EE power spectrum goes like $\te^2$. In Figure \ref{fig:tau} we compare the WMAP 7-year results to our results for the integrated $\te$ which is a consistency check on the EE power spectrum results, since there is no extra information. Comparing Figures~\ref {fig:tau} and~\ref{fig:clee} we find that the models with larger $\te$ are consistent with having larger EE power at low-$\ell$. Like the EE power spectrum, measuring $\te$ places a constraint on $\zbar$ but is unable to discern between models with small and large $\Delz$.

Here the assumption is that $\te$ is uniform in all directions, however, reionization is naturally inhomogeneous. The optical depth as a function of the distance from an observer to the CMB is  given by the equation, 

\begin{equation}
\te(l_*,\hat{{\bf{n}}}) = \sigma_{\rmn{T}} \int^{l_*}_0 n_{\rmn{e}} (\hat{{\bf{n}}},l)  dl, 
\label{eq:tau}
\end{equation}

\begin{figure}
\epsscale{1.2}
\plotone{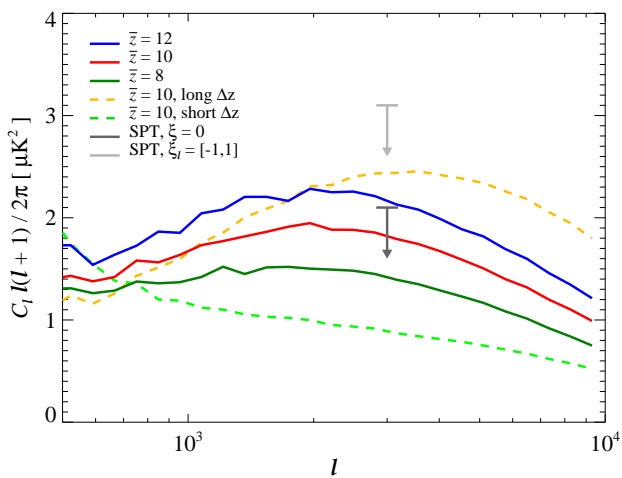}
\caption{The patchy kSZ power spectrum for various reionization models compared to SPT constraints \citep{Zahn2012}. We show five models, our fiducial model at $\bar{z} =$ 8,10, and 12 (green, red, and blue, respectively) and two extreme models of brief (orange dashed) and long (light green dashed) duration reionization at a fixed $\bar{z}=10$. The SPT constraints are illustrated by the grey arrows, with the darker grey arrow representing the constraint ignoring the tSZ-CIB cross-correlation and light including this correlation. All our models fall below the constraint that allows for tSZ-CIB correlation with an $\ell$ dependent shape and most of the models are below the even tighter constraint which ignores this correlation.}
\label{fig:kszps}
\end{figure}

\noindent where $\sigma_{\rmn{T}}$ is the Thomson cross section, $\hat{{\bf{n}}}$ is the direction normal unit vector, $dl  = c\,dt$ is the proper distance along the line of sight, $l_*$ 
is the distance to the surface of last scattering, and $n_{\rmn{e}} (\hat{{\bf{n}}},l)$ is the ionized election number density at position $l$ in the direction $\hat{{\bf{n}}}$. Previously, there was an upper limit constraint put on the RMS fluctuation in $\te$ of, at most, a few percent of the mean value $\langle\te\rangle$ using published SPT data \citep{Mort2010}. Given this upper limit the patchy $\te$ contribution to the CMB power spectrum at high-$\ell$ is negligible. In a companion paper \citep{ZR2}, we show that four point statistics of the CMB, in principle, can constrain the RMS fluctuation in $\te$ and if measured one can differentiate between models with small and large $\Delz$ and possibly break this degeneracy between $\Delz$ and $\zbar$.

\subsection{Patchy Kinetic Sunyaev Zel'dovich Power Spectrum}

The kSZ signal from patchy reionization is sensitive to the details of reionization such as the mean redshift of reionization and its duration \citep[e.g.][]{McQn2005,Zahn2005,Mess2012,Zahn2012}, which, physically, is primarily due to the relatively steep dependence on redshift of the signal strength and the dependence of ionization bubble size on duration.  In this section, we investigate how the patchy kSZ power spectrum depends on our model parameters and we compare our results against previous work and observational constraints. We provide a simple scaling relation for the patchy kSZ power at $\ell = 3000$ as a function of $\bar{z}$ and $\Delz$, which makes model fitting of observational spectra trivial.

We construct patchy kSZ maps by raytracing through the past light cone ($5.5<z<20$). The temperature distortion along each line of sight is given by

\begin{equation}
\frac{\Delta \rmn{T}}{\rmn{T}} (\hat{{\bf{n}}} ) = \frac{\sigma_{\rmn{T}}}{c} \int_{l_{\rmn{o}}}^l e^{-\te(l,\hat{{\bf{n}}})}  n_{\rmn{e}} (l,\hat{{\bf{n}}}) \hat{{\bf{n}}} \cdot {\bf{v}} dl,
\end{equation}

\noindent where ${\bf{v}}$ is the peculiar velocity, $\te(\hat{{\bf{n}}})$ is from Eq. \ref{eq:tau}, and $l_{\rmn{o}}$ is the proper distance at $z$ = 5.5. We make flat sky maps that are approximately $15^{\circ} \times 15^{\circ}$, where the angular size is determined by the N-body simulation box length of 2 Gpc/$h$ over the comoving distance out to $z=20$. Since the box length (L = 2 Gpc/$h$) of the N-body simulations is approximately equal to the comoving distance between $z=6$ and $z=20$, we cycle through the projection direction coordinates approximately once when making the maps. For each choice of parameters we make three maps along three independent axes. In Fig. \ref{fig:kszmap} we show the patchy kSZ Compton-y maps for our fiducial model and the two extreme models for the same projection direction. From these maps it is obvious how the duration of ionization affects the patchy kSZ. The models of reionization with large $\Delz$ have more small scale structure than models with small $\Delz$.

Using the flat sky approximation we calculate the power spectrum from the patchy kSZ maps and average over each projection axis for a given parameterization. Figure \ref{fig:kszps} shows the patchy kSZ power spectrum for various models of reionization from the $15^{\circ} \times 15^{\circ}$ patches. We find that increasing $\zbar$ with a fixed $\Delz$ increases the overall amplitude of the patchy kSZ power spectrum, but has little effect on the shape since by $\zbar$ does not affect the shape of $\xe(z)$ or $\Delz$. Altering $\Delz$ at fixed $\zbar$ dramatically changes both the amplitude and shape of the patchy kSZ power spectrum, since $\Delz$ affects the correlation between ionization regions \citep{ZR1}.
 Models with smaller $\Delz$ have large ionization regions, and more power at smaller $\ell$ compared to larger $\Delz$ models where the power peaks at larger $\ell$ due to the smaller coherent ionized regions. These results are qualitatively similar to \citet{Mess2012,Zahn2012}.
The strong shape and amplitude dependencies of the patchy kSZ power spectrum on $\Delz$ illustrates that if there are constraints on $\zbar$ from the EE power spectrum and $\te$, then a measurement of the patchy kSZ power will break the degeneracy between $\bar{z}$ and $\Delz$, which was already shown using SPT results \citep{Zahn2012}.

\begin{figure}
\epsscale{1.2}
\plotone{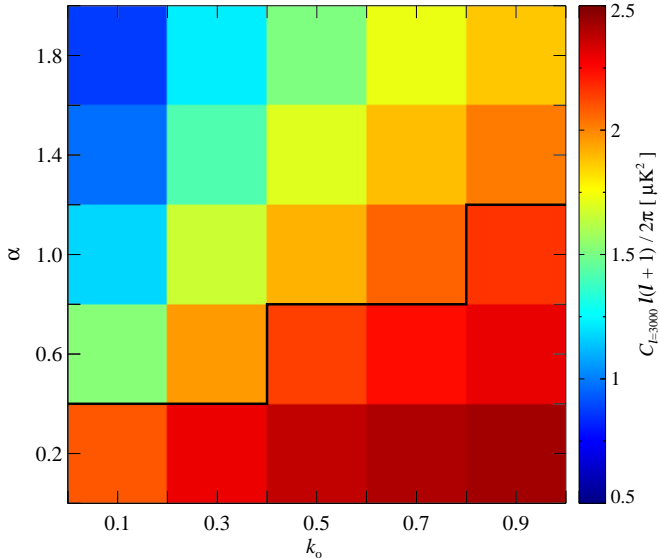}
\caption{Parameter study of the patchy kSZ power spectrum at $\ell = 3000$ for varying $\kb$ and $\al$ with a fixed $\zbar = 10$. The black line indicates the upper limit from SPT \citep{Zahn2012} with $\xi = 0$ and all models above that are consistent with this constraint. The bottom right corner has the largest $\Delz$ and the top left corner has the shortest $\Delz$.}
\label{fig:kszgrid}
\end{figure}

In Figure \ref{fig:kszps}, our patchy kSZ power spectra are compared against the SPT upper limits \citep{Zahn2012}, which accounted for the homogenous kSZ contribution using a model from \citet{Shaw2012}. Here the tightest constraint from SPT does not account for any correlation, $\xi$, between thermal SZ and the cosmic infrared background (CIB), which they measured at $\ell = 3000$ to be $\xi_{\ell=3000} = -0.18\pm 0.2$ \citep{Reic2012}. The other two constraints account for a non-zero $\xi$, where $\xi$ is given a uniform prior from -1 to 1 and either includes $\ell$ dependent shape constraint or does not. We find that our fiducial model is consistent with all SPT upper limits regardless of tSZ-CIB correlation treatment. Not all our models are consistent with SPT upper limits, for example models where we increase $\zbar$ or $\Delz$. Figure \ref{fig:kszgrid} shows our parameter space study of $\al$ and $\kb$ for the patchy kSZ power at $\ell = 3000$ with a fixed $\bar{z} = 10$. The amplitudes of the patchy kSZ power in Fig.~\ref{fig:kszgrid} range from  0.87 -  2.42 $\mu$K$^2$ and some models with large $\Delz$ (i.e. low $\al$ and large $\kb$) do not fall below the SPT constraints that exclude the tSZ-CIB correlation (cf. Fig.~\ref{fig:kszps}).

\begin{figure}
\epsscale{1.2}
\plotone{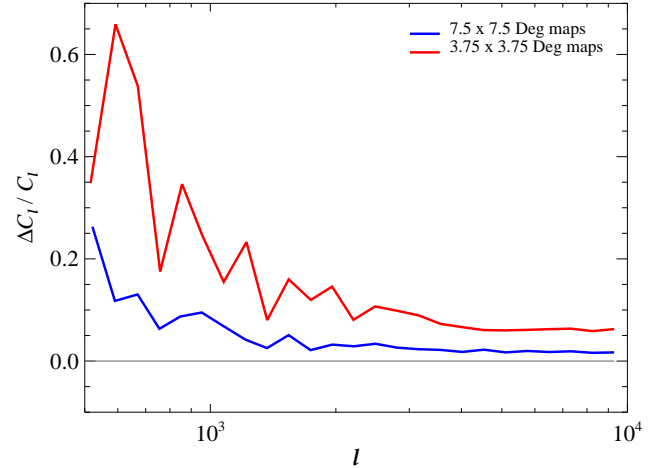}
\caption{The percent difference between the average patchy kSZ power spectra from small cut out maps compared against our $15^{\circ} \times 15^{\circ}$ maps using the fiducial parameters with a $\zbar = 10$. The average power from the $7.5^{\circ} \times 7.5^{\circ}$ cut out maps is blue line and the average power from the $3.75^{\circ} \times 3.75^{\circ}$ cut out maps is the red line. All these small maps have more power than our $15^{\circ} \times 15^{\circ}$ maps.}
\label{fig:ksz_comp}
\end{figure}

\begin{figure*}
  \resizebox{0.5\hsize}{!}{\includegraphics{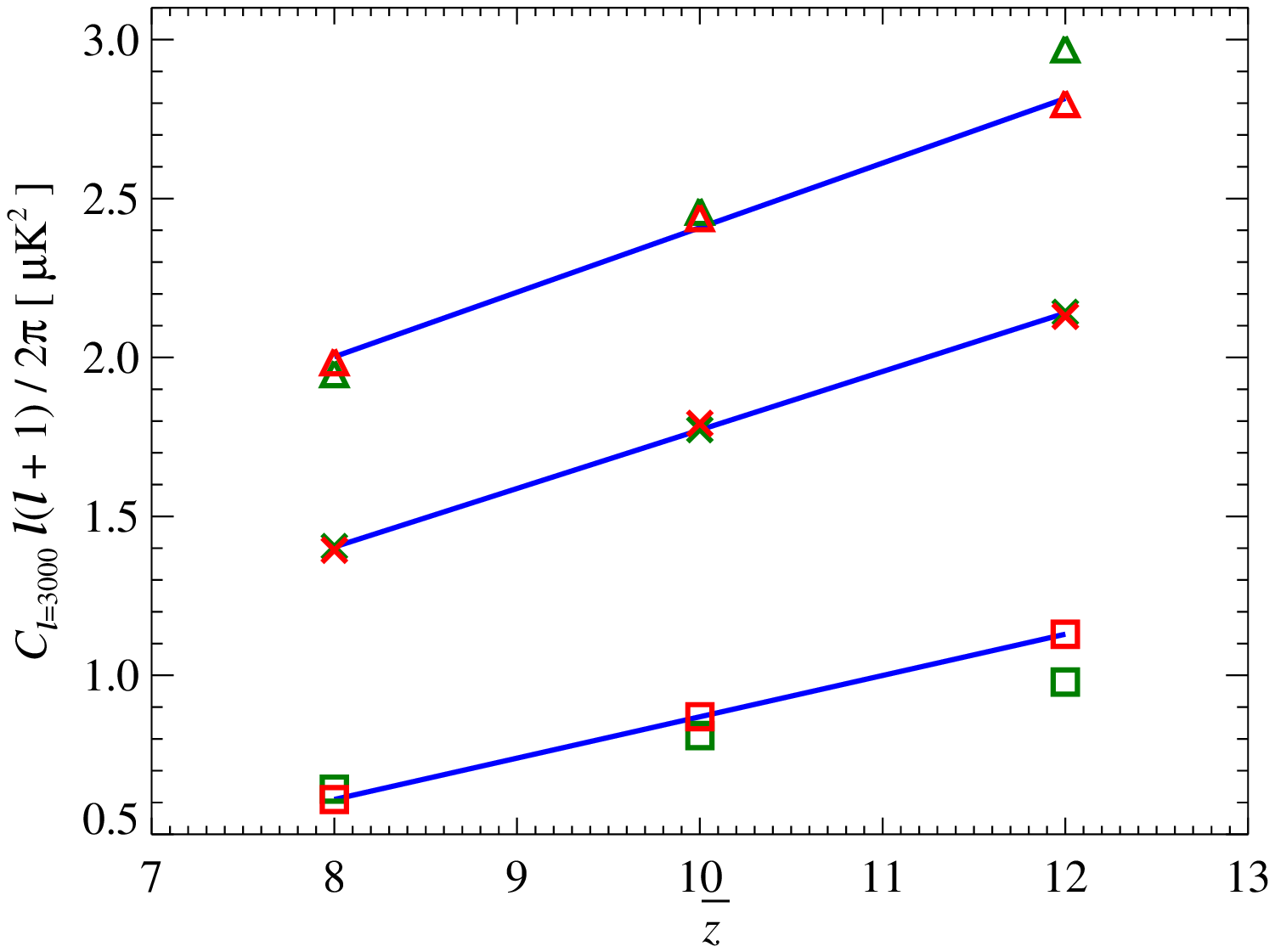}}%
  \resizebox{0.5\hsize}{!}{\includegraphics{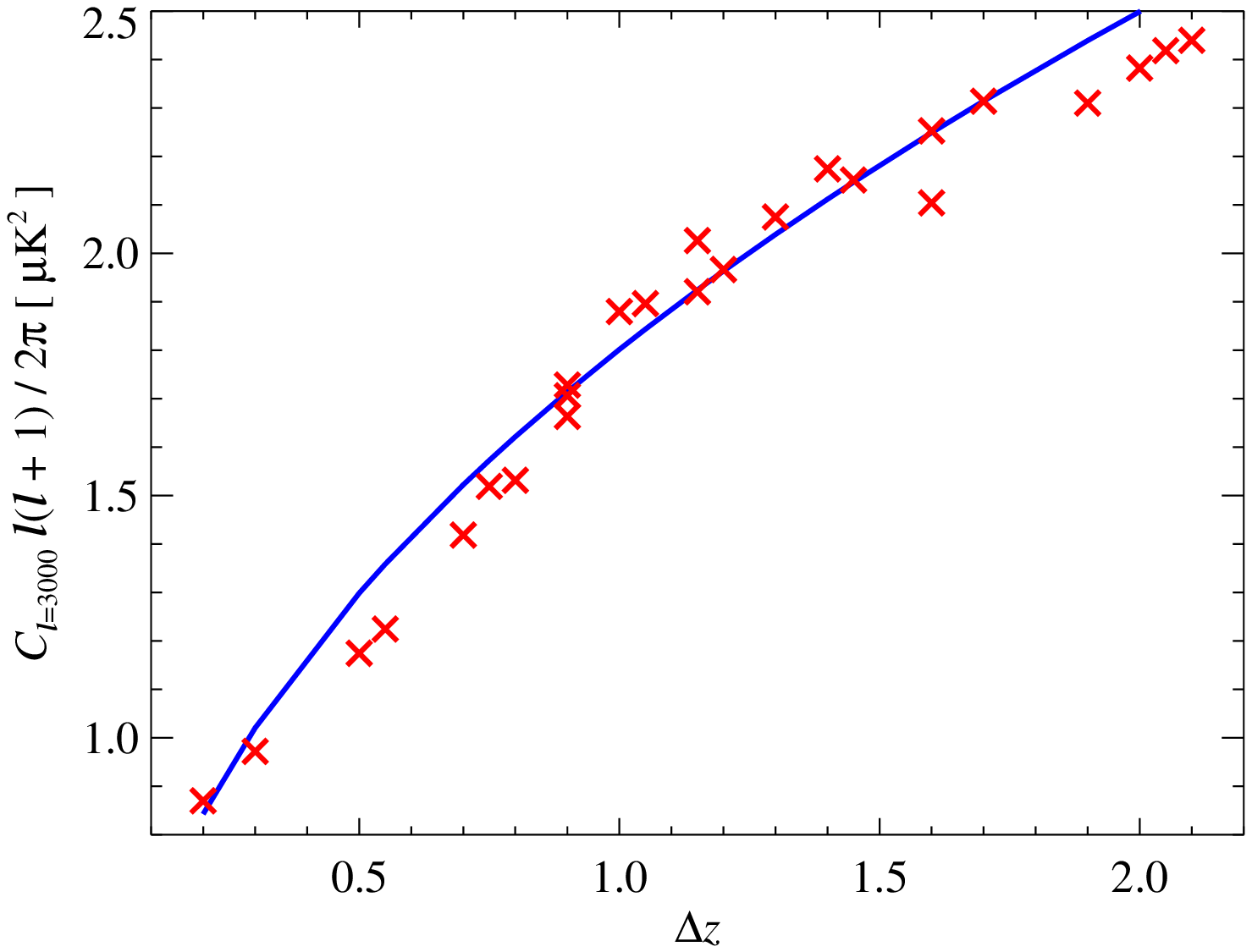}}\\                                                                                                 
  \caption{The scaling relations between the patchy kSZ amplitude at $\ell = 3000$ and $\Delz$ and $\zbar$, here the red symbols are the values calculated from the maps, the blue lines are the best fit, and green symbols are the values from the combined scaling law (cf. Eq.~\ref{eq:fit_fun}). Left: The patchy kSZ amplitude at $\ell = 3000$ as a function of $\zbar$ scales like $D_{\ell=3000}^{\rmn{kSZ}} \propto \zbar$. The squares, crosses, and triangles represent $\Delz =$ 0.2, 1.05, and 2.10, respectively. Right: The patchy kSZ amplitude at $\ell = 3000$ as a function of $\Delz$ scales like $D_{\ell=3000}^{\rmn{kSZ}} \propto \Delz^{0.5}$, here $\zbar =10$. This combined scaling relation can be used to constrain EoR in secondary parameter fitting of the high-$\ell$ CMB measurements.}  
\label{fig:kszdelz}
\end{figure*}

The kSZ power spectra that our model produces are lower than previous predictions from semi-analytic models and simulations \citep[e.g][]{McQn2005,Zahn2005,iliv2007,Zahn2012,Mess2012}. Some of these earlier works used much smaller volumes to calculate their patchy kSZ signal \citep[e.g.][]{McQn2005,Zahn2005,iliv2007}. The later works either used models with restricted parameter space \citep{Zahn2012} to fit the data, or used the Limber approximation to calculate the patchy kSZ  from 3D ionization fields \citep{Mess2012}. Testing these assumptions is left for future work.

We calculated the patchy kSZ using the flat sky approximation on $15^{\circ} \times 15^{\circ}$ maps. There are large structures in these maps that would not be captured by small scale maps (cf. Fig.~\ref{fig:kszmap}). We checked how calculating the patchy kSZ power spectrum on smaller maps would bias the patchy kSZ signal by dividing our maps into four $7.5^{\circ} \times 7.5^{\circ}$ and sixteen $3.75^{\circ} \times 3.75^{\circ}$ maps (totaling 12 and 48 maps, respectively) and calculating the spectra for each new map and averaging them together. In Figure ~\ref{fig:ksz_comp}, we show that compared to our original kSZ power spectrum the average power spectrum of the small maps had more power. At $\ell =3000$ the fractional differences are $\sim2$\% and  $\sim10$\% for the $7.5^{\circ} \times 7.5^{\circ}$ and $3.75^{\circ} \times 3.75^{\circ}$ maps, respectively. In the flat sky approximation small area maps are susceptible to erroneously produce more power on all scales since there are large scale features that these small maps do not capture. 

We found that the dependence of the amplitude of the patchy kSZ power spectrum at $\ell = 3000$ on $\zbar$ and $\Delz$ can be represented by simple scaling laws. Figure \ref{fig:kszdelz} illustrates these dependencies of $D_{\ell=3000}^{\rmn{kSZ}} \equiv C_{\ell=3000}^{\rmn{kSZ}} \ell (\ell +1) / (2\pi)$, which is linear for $\zbar$ and a power law for $\Delz$. When fitting for the scaling laws we use a nonlinear least-squares method where each value is weighted by the inverse of the variance of the three different projections and treat the dependence of $D_{\ell=3000}^{\rmn{kSZ}}$ on $\zbar$ and $\Delz$ as separable functions. The scaling laws are constrained to be:

\begin{equation}
D_{\ell=3000}^{\rmn{kSZ}}= 1.80\mu\rmn{K}^2   \left[1.12\left(\frac{1 + \zbar}{11}\right) - 0.14\right],
\label{eq:zbar_cl}
\end{equation}

\noindent for a fixed $\Delz = 1.05$ and

\begin{equation}
D_{\ell=3000}^{\rmn{kSZ}}= 1.80\mu\rmn{K}^2 \left(\frac{\Delz}{1.05}\right)^{0.47},
\label{eq:delz_cl}
\end{equation}

\noindent for a fixed $\zbar = 10$. The combination of Eq. \ref{eq:zbar_cl} and \ref{eq:delz_cl}

\begin{equation}
D_{\ell=3000}^{\rmn{kSZ}} \simeq 2.02 \mu\rmn{K}^2  \left[\left(\frac{1 + \zbar}{11}\right) - 0.12\right] \left(\frac{\Delz}{1.05}\right)^{0.47}
\label{eq:fit_fun}
\end{equation}

The predicted $D_{\ell=3000}^{\rmn{kSZ}}$ from Eq. \ref{eq:fit_fun} compares well to the results from our model (cf. Fig.~\ref{fig:kszdelz}). The slight deviations are seen in the variation of the map power spectrum values about the fit (right panel Fig. \ref{fig:kszdelz}) and these deviations are found at the extreme ends of parameter space. 
Using Eq. \ref{eq:fit_fun} we find a lower limit of $D_{\ell=3000}^{\rmn{kSZ}} \gtrsim 0.4 \mu$K$^2$ by taking the $2\sigma$ lower confidence interval on $\zbar = 8.1$ from WAMP7 and the lower limit on $\Delz \gtrsim 0.07$ from EDGES. Here we have converted the EDGES definition of $\Delz$, which assumes a functional form of hyperbolic tangent for $\xe(z)$ to our definition. This scaling law provides a simple way to place model dependent constraints on $\zbar$ or $\Delz$ by including it when fitting high-$\ell$ CMB power spectra measurements into the secondary models used. However this requires additional measurements, for example of the EE power spectrum, to break the degeneracies between $\Delz$ and $\zbar$ that occurs when just using patchy kSZ measurements. 

\section{Future Constraints}

Constraints on $\zbar$ and $\Delz$ will tighten as the precision increases on measurements of the low-$\ell$ EE polarization and high-$\ell$ temperature power spectra.
We forecast how well these future precision measurements of $\te$ and $D_{\ell=3000}$ will constrain $\zbar$ and $\Delz$ by constructing a likelihood surface from a $\chi^2$ grid of $\zbar$ and $\Delz$ around our fiducial model. The $\chi^2$ grid is calculated following 

\begin{equation}
\chi^2 = \left[\frac{\te - \te_{\rmn{fid}}}{\sigma_{\te}} \right]^2 + \left[\frac{D_{\ell=3000} - D_{\rmn{\ell=3000,fid}}}{\sigma_{D_{\ell=3000}}}\right]^2,
\end{equation}

\noindent here $\te_{\rmn{fid}}$ and $D_{\rmn{\ell=3000,fid}}$ are the values for $\te$ and $D_{\rmn{\ell=3000}}$ from the fiducial model, $\sigma_{\te}$ is the forecasted error bar for Planck or CMBpol on $\te$, and $\sigma_{D_{\ell=3000}}$ is the hypothetical error bar for ACT-pol and SPT-pol on $D_{\rmn{\ell=3000}}$. 
Using our fiducial value for $\te$ we estimate that Planck will measure $\te \pm 0.004$ \citep[$\sim5$\% error;][]{PlanckBB} and CMBpol will measure $\te \pm 0.002$ a factor of 2 better \citep{Zald2008}. There is still no detection of the patchy kSZ power spectrum, only upper limits \citep{Reic2012,Zahn2012}. A detection of the patchy kSZ power spectrum will depend upon the ability to properly model contributions from the thermal SZ power spectrum \citep[which depends on uncertain intracluster medium astrophysics e.g.][]{Batt2010,Shaw2010,Trac2011a,Batt2012}, the homogeneous kSZ \citep[e.g.][]{Ostr1986,Jaff1998,Ma2002,Zhan2004,Shaw2012}, the thermal SZ  - CIB cross spectrum \citep{Reic2012,Addi2012,Zahn2012}, and the infrared and radio sources \citep[see][for these models]{Dunk2011,Reic2012}. We choose two hypothetical error bar values for ACT-pol and SPT-pol measurements of $D_{\rmn{\ell=3000}}$.

\begin{figure}
\epsscale{1.2}
\plotone{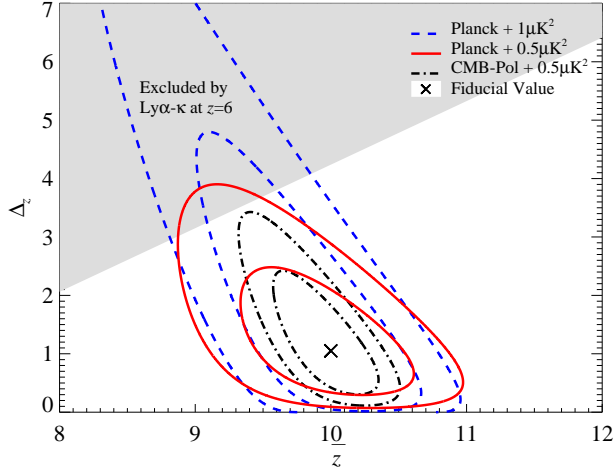}
\caption{Projected constraints on $\zbar$ and $\Delz$ for our fiducial model of EoR given the estimated error bars for Planck and CMBpol, and optimistic hypothetical error bars for ACT-pol and SPT-pol. The 1$\sigma$ (inner ellipses) and 2$\sigma$ (outer ellipses) constraints for Planck projected error bars on $\te$ with 1$\mu$K$^2$ (blue dashed lines) and 0.5$\mu$K$^2$ (red lines) estimated error bars for ACT-pol and SPT-pol on $D_{\ell=3000}$ Patchy kSZ measurements. The black dot dashed lines are the 1$\sigma$ and 2$\sigma$ constraints for CMBpol projected error bars on $\te$ and  0.5$\mu$K$^2$ estimated error bars for ACT-pol and SPT-pol on $D_{\ell=3000}$ Patchy kSZ measurements. The grey region across the top is excluded by the zero transmission of rest-frame $\LAF$ flux at $z\gtrsim6$ in spectra of quasars \citep[$\LAF$-$\kappa$;][]{Fan2006a}.}
\label{fig:const}
\end{figure}

The likelihood surface is
$\Lik \propto e^{-\chi^2 / 2}$,
and the $1\sigma$ and $2\sigma$ contours are the 68\% and 95\% probability surface. Figure \ref{fig:const} show the forecasted constraints on $\zbar$ and $\Delz$ from future measurements. Results from Planck and a detection of the patchy kSZ power at $\ell=3000$ will constrain $\zbar$ to $\sim 5$\%. The value of $\Delz$ begins to be constrained when we combine the upper limit from the opacity of the $\LAF$ forest ($\LAF$-$\kappa$) at $z=6$ in quasar spectra. Here the upper limit from the opacity of the $\LAF$ forest is derived by converting our $\Delz$ to $z(\xe = 50\%) - z(\xe = 90\%)$. Figure \ref{fig:const} illustrates that an experiment like CMBpol will tighten these constraints tremendously. While we focused our analysis on $\te$ and $D_{\rmn{\ell=3000}}$ it is possible that the low-$\ell$ EE power spectrum from CMBpol could provide constraints on $\Delz$ as well, which would make our forecasted constraints even better.  

\section{Conclusions}
\label{sec:con}

Using a new semi-analytic model for constructing the reionization-redshift field from any density field, we made predictions for the low-$\ell$ polarization power spectrum and the high-$\ell$ temperature spectrum measurements of the CMB. We demonstrated that combining measurements of the EE power spectrum with the patchy kSZ amplitude at $\ell =3000$ constrains both the mean reionization-redshift and the duration of reionization. The measured EE power spectrum from WMAP and the predicted improved spectrum from Planck will constrain $\zbar$, but cannot discern between models for reionization with extreme durations.

The shape and the amplitude of the patchy kSZ power spectrum depend on both the duration and the mean redshift of reionization. At $\ell = 3000$, where the amplitude of the kSZ power spectrum is currently constrained, we find the patchy kSZ power at $\ell=3000$ ranges from 0.87 - 2.42 $\mu$K$^2$ with a $\zbar=10$ (this $\zbar$ matches the current WMAP 7-year constraints). The largest kSZ signals correspond to long duration reionization models. We found a simple scaling law for the patchy kSZ power spectrum amplitude at $\ell = 3000$ as a function of $\zbar$ and $\Delz$, which makes model fitting to observed spectra trivial. Using this scaling law and constraints from WMAP on the $\zbar$ and the lower limit from EDGES we place a lower limit on the patchy kSZ amplitude at $\ell=3000$ of $\sim 0.4\ \mu$K$^2$.

The amplitudes we find for the patchy kSZ power spectra are lower than previous model predictions. The differences between our work and the earlier work on the patchy kSZ signal \citep{McQn2005,Zahn2005,iliv2007} are that previous work used much smaller volumes to calculate this signal. When small volumes are used to calculate the patchy kSZ signal, we show that the amplitudes of the power spectra are biased to larger values. 

Many of our reionization models are consistent with the tightest upper limit constraint from SPT \citep{Zahn2012}, including our fiducial model, which has a $\zbar = 10$ and $\te = 0.085$ to be consistent with the current WMAP 7-year constraints. This consistency is achieved without the need to invoke more exotic or unphysical models for reionization that were necessary for the previous models of the patchy kSZ in order to fit within these constraints. In the event that the measured values of $\te$ and $\zbar$ decrease, the patchy kSZ power for all models would decrease further.

It is clear that the future measurements from the Planck satellite of the EE power spectrum will tightly constrain the mean reionization-redshift and CMBpol has the potential to do even better. This leaves measurements of the patchy kSZ power spectrum from high resolution CMB observations to constrain the duration of the EoR. The current upper limits of the patchy kSZ amplitude at $\ell = 3000$ from SPT range from 2.1-4.9 $\mu$K$^2$ depending on the assumptions made about the correlation between the tSZ and CIB, but future detections are projected to be $\sim 1\ \mu$K$^2$ \citep{Reic2012}. In order to measure the patchy kSZ, we first need to understand $\xi$ \citep{Reic2012,Addi2012,Zahn2012}, have a good understanding of the contributions from the homogenous kSZ \citep[e.g.][]{Ostr1986,Jaff1998,Ma2002,Zhan2004,Shaw2012}, and how the astrophysical uncertainties of the homogenous kSZ models. A measurement of the patchy kSZ will tighten the constraints on $\zbar$ and $\Delz$ greatly.

Combining these CMB constraints with neutral hydrogen measurements \citep[e.g][]{ZR4}, such as redshifted 21cm signal that originates from the hyperfine transition of neutral hydrogen \citep[e.g.][]{Scot1990,Shav1999,Zald2004}, will constrain the mean redshift and duration of reionization further, or provide new issues for reionization models to tackle.

\acknowledgments

N.B. and A.N. are supported by a McWilliams Center for Cosmology Postdoctoral Fellowship made possible by Bruce and Astrid McWilliams Center for Cosmology. We thank  Graeme Addison, Paul La Plante, Christian Reichardt, and Jonathan Sievers for useful discussions. H.T. is supported in part by NSF grant AST-1109730. R.C. is supported in part by NSF grant AST-1108700 and NASA grant NNX12AF91G. A.L. is supported in part by NSF grant AST-0907890 and NASA grants NNX08AL43G and NNA09DB30A. The simulations were performed at the Pittsburgh Supercomputing Center (PSC) and the Princeton Institute for Computational Science and Engineering (PICSciE). We thank Roberto Gomez and Rick Costa at the PSC and Bill Wichser at PICSciE for invaluable help with computing.

\bibliography{nab_PR}
\bibliographystyle{apj}

\end{document}